\documentclass{aa}  
\usepackage{graphicx}
\usepackage{txfonts}
\usepackage{longtable}
\usepackage{natbib}
\bibpunct{(}{)}{;}{a}{}{,}
\begin{document}
   \title{Lucky Imaging of M subdwarfs\thanks{Based on observations collected at the 
   Centro Astron\'omico Hispano Alem\'an (CAHA) at Calar Alto, operated jointly by the 
   Max-Planck Institut f\"ur Astronomie and the Instituto de Astrof\'isica de Andaluc\'ia (CSIC)}}

   \subtitle{}

   \author{N. Lodieu \inst{1}
          \and
          M. R. Zapatero Osorio \inst{1}
          \and
          E. L. Mart\'{\i}n \inst{1}
          }

   \offprints{N. Lodieu}

   \institute{Instituto de Astrof\'isica de Canarias, V\'ia L\'actea s/n,
                 E-38200 La Laguna, Tenerife, Spain\\
              \email{nlodieu@iac.es,mosorio@iac.es,ege@iac.es}
             }

   \date{\today{}; \today{}}

 
  \abstract
   {The knowledge of the binary properties of metal-poor and 
solar-metallicity stars can shed light on the potential differences 
between the formation processes responsible for both types of objects.}
   {The aim of the project is to determine the binary properties (separation,
mass ratio, frequency of companions) for M subdwarfs, the low-metallicity
counterparts of field M dwarfs, and investigate any potential differences
between both populations.}
   {We have obtained high-resolution imaging in the optical for a sample
of 24 early-M subdwarfs and nine extreme subdwarfs with the ``Lucky
Imaging'' technique using the AstraLux instrument on the Calar Alto
2.2-m telescope.}
   {We are sensitive to companions at separations larger than 0.1 arcsec
and differences of $\sim$2 magnitudes at 0.1 arcsec and $\sim$5 mag
at 1 arcsec. We have found no companion around the 24 subdwarfs under 
study and one close binary out of nine extreme subdwarfs. A second
image of LHS\,182 taken three months later with the same instrument
confirms the common proper motion of the binary separated by about 
0.7 arcsec. Moreover, we do not confirm the common proper motion of the 
faint source reported by Riaz and collaborators at $\sim$2 arcsec from 
LHS\,1074\@.}
   {We derive a binary frequency of 3$\pm$3\% for M subdwarfs 
from our sample of 33 objects for separations larger than about five 
astronomical units. Adding to our sample the additional 28 metal-poor 
early-M dwarfs observed with the Hubble Space Telescope by
Riaz and collaborators, we infer a binary fraction of 3.7$\pm$2.6\%
(with a 1$\sigma$ confidence limit), significantly lower than the 
fraction of resolved binary M dwarfs ($\sim$20\%) 
over the same mass and separation ranges. This result suggests a sharp 
cut-off in the multiplicity fraction from G to M subdwarfs, indicating
that the metallicity plays a role at lower masses and/or an environmental
effect governing the formation of metal-poor M dwarfs compared to their
metallicity counterparts.
}

   \maketitle
%

%
%
\section{Introduction}
\label{subdw_CAHA:Intro}

Cool subdwarfs are metal-poor dwarfs which appear less luminous
than their solar-metallicity counterparts because their atmospheres
are deficient in metals \citep{baraffe97}. They exhibit halo kinematics,
high proper motions and high heliocentric velocities \citep{gizis97a}. 
They are typically very old (10 Gyr or more) and likely belong to the 
first generations of stars in the Galaxy.
The adopted classification for M subdwarfs (sdM) and extreme subdwarfs
(esdM) originally proposed by \citet{gizis97a} has recently been revised 
by \citet{lepine07c}. A new class, the ultra-subdwarfs (usdM), 
has been added to the sdM and esdM originally proposed.
The new scheme is based on a parameter, $\zeta_{\rm TiO/CaH}$, a proxy
for the weakening of the strength of the TiO band as a function of 
metallicity. An alternative classification scheme has been proposed by
\citet{jao08} by comparing model grids with optical
spectra to characterise the spectral energy distribution of subdwarfs
by three parameters: temperature, gravity, and metallicity.

The binary frequency of solar-metallicity dwarfs seem to decrease with
the primary mass. Results from visual, speckle, and
spectroscopic studies of massive O and B stars (isolated or in clusters)
have revealed that a significant number are in multiple systems
\citep[59\%;][]{mason98}. At lower masses,
\citet{duquennoy91} have inferred a multiplicity of
57$\pm$9\% for a sample of 164 solar-metallicity stars. These systems 
have mass ratios larger than 0.1 and harbour a wide range of separations.
M dwarf multiples are as frequent as 42$\pm$9\% and have binary 
properties such as mass ratio and separation similar to G dwarfs
\citep{henry90a,fischer92}. In the substellar regime (total mass 
less than 0.1 M$_{\odot}$), the multiplicity seems lower (10--20\%) 
with a preference for small separations ($\sim$4--8 AU) and 
equal-mass systems \citep{martin03,burgasser07a}.

Population II stars are old (10\,Gyr or more), implying that any
object more massive than a solar-metallicity star has evolved off the
main-sequence into a white dwarf.
Independent studies of subdwarfs drawn from the Carney-Latham 
catalogue \citep{carney94} across a large separation range favour
similar binary fractions between metal-poor and solar-metallicity 
stars with G spectral types.
This conclusion is valid for spectroscopic binaries 
\citep[$<$3 AU;][]{stryker85,latham02}, separations from a few to tenths 
of astronomical units \citep[AU;][]{koehler00b,zinnecker04}, and wide
binaries \citep{allen00,zapatero04a}. In the M dwarf regime, the sample
imaged at high resolution is much smaller. 
\citet{riaz08a} expanded the {\it{Hubble Space Telescope}} ({\it{HST}}) 
high-resolution imaging by \citet{gizis00c} and derived an upper limit 
on the binary frequency of 7\% (3.6$\pm$3.6\%; 1$\sigma$ limit) from
a sample of 28 M subdwarfs. An independant sample of 18 sdM 
were observed with the Lick adaptive optics Laser Guide Star system 
by \citet{lepine07b} and one sdK7.5, LHS\,1589, was resolved into a 
close binary system, yielding a binary fraction of 5.6$\pm$5.6\%.

The fraction of stars hosting planets is larger with higher metallicity. 
The average metallicity of a volume-limited sample of stars with planets 
and non binary stars (that have been specifically searched for planets) peaks 
at about [Fe/H] = $+$0.1 and $-$0.1, respectively \citep{santos05a,bond06}.
The frequency of metal-poor stars with planets is of the order of
five percent or less, increasing to 30\% for stars with metallicity
of $+$0.25 \citep{santos01a,santos05a}. This trend is now widely accepted
and is not the result of observational effects (e.g., lack of metal-poor
stars in the solar neighbourhood or weakness of the absorption lines at 
lower metallicity). Moreover, there might be a trend towards low-mass
planets with short periods orbiting low metallicity stars
\citep{santos03a,pinotti05}.

What is the role of metallicity in the binary properties of low-mass
stars? Our aim is to provide a first response to this question 
and bridge the gap between the high-resolution imaging surveys of 
metal-poor G stars and radial velocity studies of extrasolar planets 
around solar-metallicity stars.
In this paper we present high-resolution $z$-band imaging for a sample
of 24 sdM and nine esdM obtained with the AstraLux camera on the Calar 
Alto 2.2-m telescope. In Sect.\ \ref{subdw_CAHA:sample} we describe the 
selection of the sample. In Sect.\ \ref{subdw_CAHA:Obs} we present 
additional photometric and astrometric measurements for the possible 
faint companion to LHS\,1074 detected by \citet{riaz08a} as well as
the observations, data reduction, analysis of the AstraLux data.
In Sect.\ \ref{subdw_CAHA:BF} we discuss the binary frequency of early-M 
subdwarfs and compare our results to previous surveys dedicated to the 
multiplicity of subdwarfs.

%
%
%
\section{Sample selection}
\label{subdw_CAHA:sample}

We have selected the brightest known subdwarfs with spectral types
derived from optical spectroscopy by various authors 
(Table \ref{tab_subdw_CAHA:sample_observed}). There are 24 subdwarfs
and nine extreme subdwarfs with metallicites [m/H] of approximately $-$1.2
and $-$2.0, respectively \citep{gizis97a}. In total, our sample contains 33 
metal-poor early-M with spectral types between M0 and M5\@.
They are typically brighter than $I \sim$ 14 mag. Because they are bright, 
they are usually among the closest M subdwarfs with distances less than 
100 pc (Table \ref{tab_subdw_CAHA:sample_observed}). However, only four
subdwarfs, identified in the SUPERBLINK catalogue \citep{lepine02}, are
beyond 50 pc. Theoretical models predict masses around 
$\sim$0.1--0.4 M$_{\odot}$ and $\sim$0.09--0.2 M$_{\odot}$ for early-M 
subdwarfs and extreme subdwarfs at an age of 10 Gyr \citep{baraffe97}, 
respectively, compared to $\sim$0.13--0.6 M$_{\odot}$ for solar-metallicity 
dwarfs of similar effective temperatures \citep{baraffe98}.

Our target list contains seven sources common to a similar survey conducted 
by \citet{riaz08a} with the {\it{HST}}, including LHS\,491, LHS\,320, 
LHS\,3409, LHS\,364, LHS\,536, LHS\,161, and LHS\,491 \citep{luyten79}.
In addition, we have included the LHS\,1589AB system
\citep{lepine07b} to test the performance of AstraLux and check the
common proper motion.

Moreover, we obtained a second imaging epoch with the Long-slit 
Intermediate Resolution Infrared Spectrograph \citep[LIRIS;][]{manchado98} 
on the William Herschel Telescope (WHT) of the sdM6 LHS\,1074 
to investigate the common proper motion of the faint source
located two arcsec away from LHS\,1074 \citep{riaz08a}. Both the
primary and the candidate are beyond the capabilities of AstraLux.
The astrometric study of LHS\,1074 is mandatory for a proper statistical 
analysis of the binary fraction among the low-mass metal-poor stars in
the combined surveys by \citet{riaz08a} and ours. 
Sect.\ \ref{subdw_CAHA:lhs1074} is dedicated to this object.

%
%
\section{Photometric observations}
\label{subdw_CAHA:Obs}
%

%
%
\subsection{WHT/LIRIS observations of LHS\,1074}
\label{subdw_CAHA:lhs1074}

In this section we discuss the binarity of LHS\,1074, a sdM6.0 subdwarf
with a photometric distance estimated as $\sim$86 pc \citep{reid05} and
originally listed in the Luyten Half-Second Catalog \citep{luyten79}. 

\citet{riaz08a} imaged LHS\,1074 deeply with {\it{HST}} and found a faint 
companion at $\sim$2.1 arcsec from the primary. To confirm (or otherwise)
the common proper motion of LHS\,1074 and the faint companion, we have 
obtained a deep $J$-band image with LIRIS \citep{manchado98} on the WHT
at the Roque de Los Muchachos Observatory on the island of La Palma in 
the Canaries.
LIRIS is equipped with a 1024$\times$1024 HAWAII detector working in the
0.8 to 2.5 micron wavelength range and has a scale of 0.25 arcsec per pixel,
translating into a field of view of 4.27 arcmin a side. LIRIS $J$-band 
observations were conducted on July 1, 2008. The total integration time 
was 4860 seconds. The final stacked image of LHS\,1074 consisted 
of 81 short individual exposures of 60 seconds. Raw near-infrared data 
were reduced in a standard way for these wavelengths, including sky 
substraction and flat-fielding.

The proper motion of LHS\,1074 is about 0.8 arcsec/yr \citep{luyten79}.
The epoch difference between the {\it{HST}} taken on 24 October 2003 and
WHT/LIRIS observations is approximately 4.7 years, long enough to determine
unambiguously the common proper motion of LHS\,1074 and the possible 
companion announced by \citet{riaz08a}.
On the LIRIS stacked image, we measured a separation of 5.45 arcsec
with an uncertainty of one pixel i.e., 0.25 arcsec and a $J$-band magnitude
of 21.47$\pm$0.15 (Table \ref{tab_subdw_CAHA:lhs1074_tab}). Both results
confirm that both objects do not form a physical pair. On the one hand, we
expect LHS\,1074 to have moved by 3.76 arcsec in 4.7 years which is
consistent within the uncertainties with the difference in the separation
measured on the {\it{HST}} and LIRIS images. On the other hand, the
expected temperature of a subdwarf six magnitudes fainter than its primary,
a sdM6 subdwarf, should be very red and substellar.
Those objects exhibit redder $z-J$ optical-to-infrared colours
\citep[typically 3--4 mag;][]{hawley02,knapp04}, inconsistent with
the photometry obtained in the $m_{775}$ (22.6 mag) and $J$ filters.

Table \ref{tab_subdw_CAHA:lhs1074_tab} lists the coordinates, photometry 
from {\it{HST}} ($I_{m_{775}}$) for LHS\,1074 (top line) and its possible 
companion (lower line). The $J$-band magnitudes are from 2MASS for LHS\,1074 
and LIRIS for the companion with an estimated uncertainty of 0.3 mag.
The respective epochs of the {\it{HST}} and LIRIS observations
are also given (column 5) with the measured separations given in arcsec
in the last column. The difference in separation matches the proper motion 
of LHS\,1074 ($\sim$0.8 arcsec/yr) within the measurement uncertainties.

%
%
%
%
\begin{table}
 \centering
  \caption{Coordinates, photometry, epochs of observations, and 
separations from the {\it{HST}} (1) and LIRIS observations (2). 
References: (1) \citet{riaz08a}; (2) this paper.
  }
 \label{tab_subdw_CAHA:lhs1074_tab}
 \begin{tabular}{c c c c c c}
 \hline
  R.A. (J2000)         &      Dec (J2000)   &   $I_{m_{775}}$   &  $J_{C}$  & Epoch  & Sep    \cr
  h:m:s                &      d:':"         &     mag           &  mag      & years  & arcsec \cr
 \hline
 00:25:51.32 & $-$07:48:09.3  &  16.07  & 14.68 & 2003.81 &  2.1  \cr
 00:25:51.31 & $-$07:48:11.4  &  22.60  & 21.47  & 2008.50 & 5.45 \cr
 \hline
 \end{tabular}
\end{table}
%

%
%
\subsection{AstraLux observations}
\label{subdw_CAHA:Obs_obs}

Observations were carried out with AstraLux, the ``Lucky Imaging'' facility 
installed on the Calar Alto 2.2-m telescope. AstraLux is equipped with a 
electron-multiplying, thinned, and back-illuminated 512$\times$512 CCD 
detector with a pixel scale of $\sim$47 milli-arcsec(mas), yielding a 
field-of-view of approximately 24 by 24 arcsec \citep{hormuth08a}.
The read-out noise is of the order of 80 electrons and the
distortion across the entire field is less than 0.5\%. Various optical
filters are available for observations, including the Cousins $I$ 
\citep{johnson53,cousins78} and the SDSS (Sloan Digital Sky Survey) 
$i$ and $z$ \citep{fukugita96}.

We have targeted a total of 24 sdM and nine esdM over three campaigns 
(Table \ref{tab_subdw_CAHA:sample_observed}): 14 \& 16 January, 04--05 June, 
and 07-10 November 2008\@. In January, the nights were clear with seeing 
around 1.5 arcsec. The night of June 04th was affected by clouds despite
a seeing in the 0.8--1.2 arcsec range. The following night was clear with 
good transparency but the seeing oscillated significantly with values up 
to 2 arcsec. In November, the conditions were significantly better than in
January and June with clear skies and seeing around 0.6--0.8 arcsec.
Therefore, we decided to repeat all the subdwarfs observable in November
to improve the depth of the images.
Typically, we obtained a total of 300 seconds total integration time divided
into 6000 exposures of 50 milli-seconds (ms) for our targets to sample 
the rapid seeing variations. However, those values were modified accordingly 
depending on the brightness of the source. The Strehl ratios are usually
higher than 10\%, except for a few targets 
(Table \ref{tab_subdw_CAHA:sample_observed}). We have selected the 10\% 
best-quality images resulting in a effective 30 seconds exposure time. 
We have achieved resolutions of $\sim$0.1 arcsec, typical for Lucky imaging
observations in a 2-m class telescope, for all the subdwarfs 
(Table \ref{tab_subdw_CAHA:sample_observed}), i.e., about one-tenth
of the natural seeing conditions at the time of the observations.

A series of dome flat fields and bias were taken before each night and 
skyflats after sunset when possible to create master bias and master
flats. We have also observed the M15 globular cluster
during each night to calculate the pixel scale and 
the orientation of the camera and derive the photometric zero-points
(applied only if a binary was found). 

%
%
\subsection{Data reduction}
\label{subdw_CAHA:Obs_DR}

The data reduction of the raw images are done on-the-fly with an automatic 
pipeline available at the telescope and distributed by the AstraLux team 
\citep{hormuth08a}. In the case of the sole binary resolved by our survey, 
the comparable brightness measured for the primary and the secondary leads
to a ``triple'' system with the automatic pipeline. Therefore, we 
re-processed the raw images by imposing the primary star as the reference
to remove this well-known effect present in Lucky Imaging images.
This was done using the "two star" option available
in the FastCam (PI Rafael Rebolo)\footnote{More details on FastCam at:
http://www.iac.es/proyecto/fastcam/} data reduction package developed 
by the Universidad Polyt\'ecnica de Cartagena. A similar result 
was obtained by Felix Hormuth (personal comm.) using a simple
``detripling'' algorithm implemented to remove a ``triple'' component when 
the primary and the secondary of a binary system have comparable brightness.
Although we have not been involved in the writing of the automatic 
pipeline, we give a short summary of the steps involved 
in the data processing. More specific information on 
the processing is extensively detailed in \citet{hormuth08a} and in the
PhD theses of Robert Tubbs and Nicholas Law\footnote{http://www.ast.cam.ac.uk/$\sim$optics/Lucky\_Web\_Site/references.htm} 
as well as on the AstraLux website\footnote{http://www.mpia.de/ASTRALUX/Publications.html}.

All the observations were taken using only one-fourth of the detector,
covering a 6 by 6 arcsec field-of-view.
Firstly, the raw $z$-band images were divided by the master flats
previously corrected for the bias. Secondly, the quality of 
the images was determined by measuring the frames with the largest flux 
in the point-spread function of the reference star. In our case, the target 
was used as the reference star and was usually (but not always) the only 
object in the field.
The next step consists of computing the shifts between each individual
frame to align all the images. Then, the frames are ranked on the basis
of the quality of their point-spread function. The 1, 2.5, 5, and 10\% 
best-quality images were aligned, and co-added using the Drizzle algorithm 
\citep{fruchter02} to provide a final pixel scale of 0.0233 mas. Finally, 
those images are saved to disk for visual inspection. The pipeline
processing is quicker than the typical exposure time used for our
targets and allowed us to obtain an $i$-band image of any potential
binary system before moving to the next target.

%
%
\begin{figure*}
    \centering
    \includegraphics[width=\linewidth]{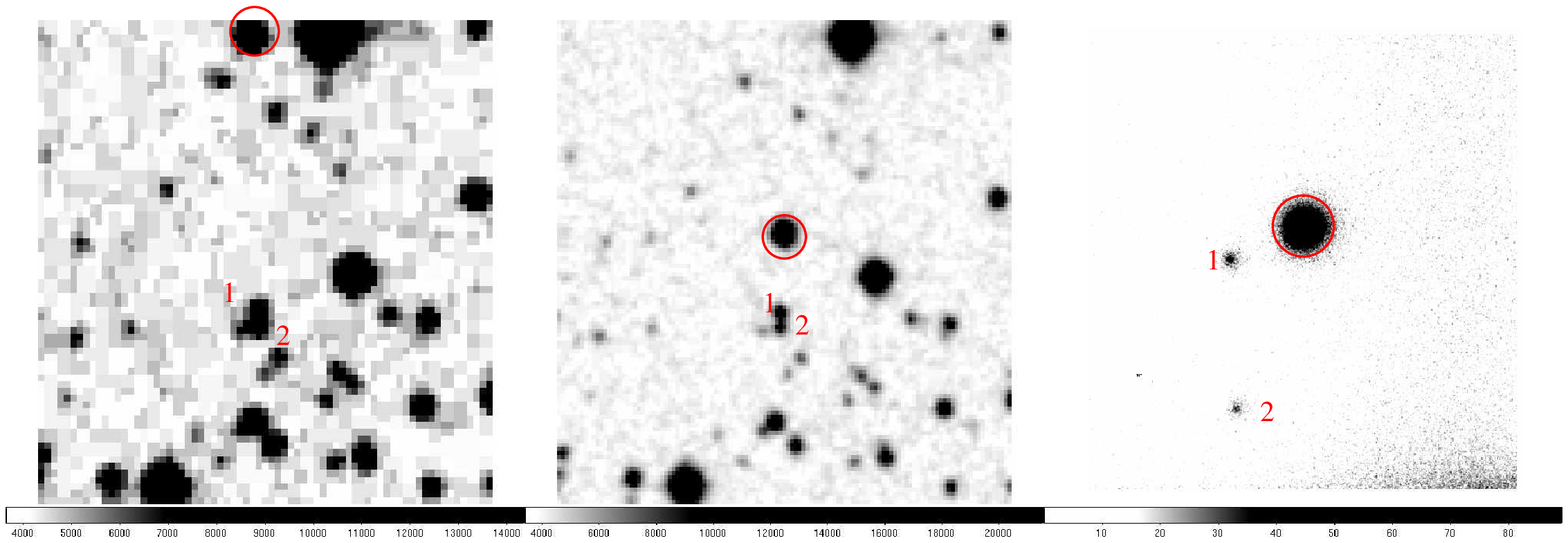}
    \includegraphics[width=\linewidth]{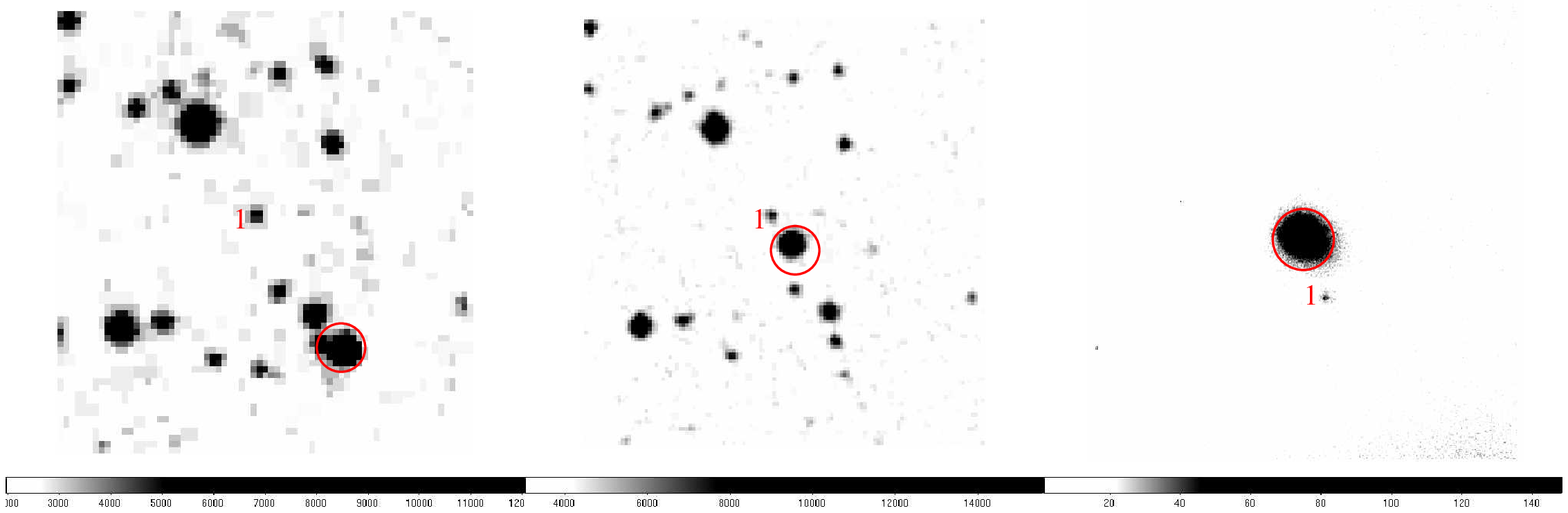}
    \caption{Images showing the motion of LHS\,489 (top) and
    LSR J2009$+$5659 over $\sim$55 years on POSS I (year 1953; left),
    POSS II (year 1990; middle), and AstraLux images (SDSSz; November 2008; 
    right).
 }
    \label{fig_subdw_CAHA:images_motion}
 \end{figure*}
%

%
%
\subsection{Analysis}
\label{subdw_CAHA:analysis}

\begin{figure*}
   \centering
   \includegraphics[width=\linewidth]{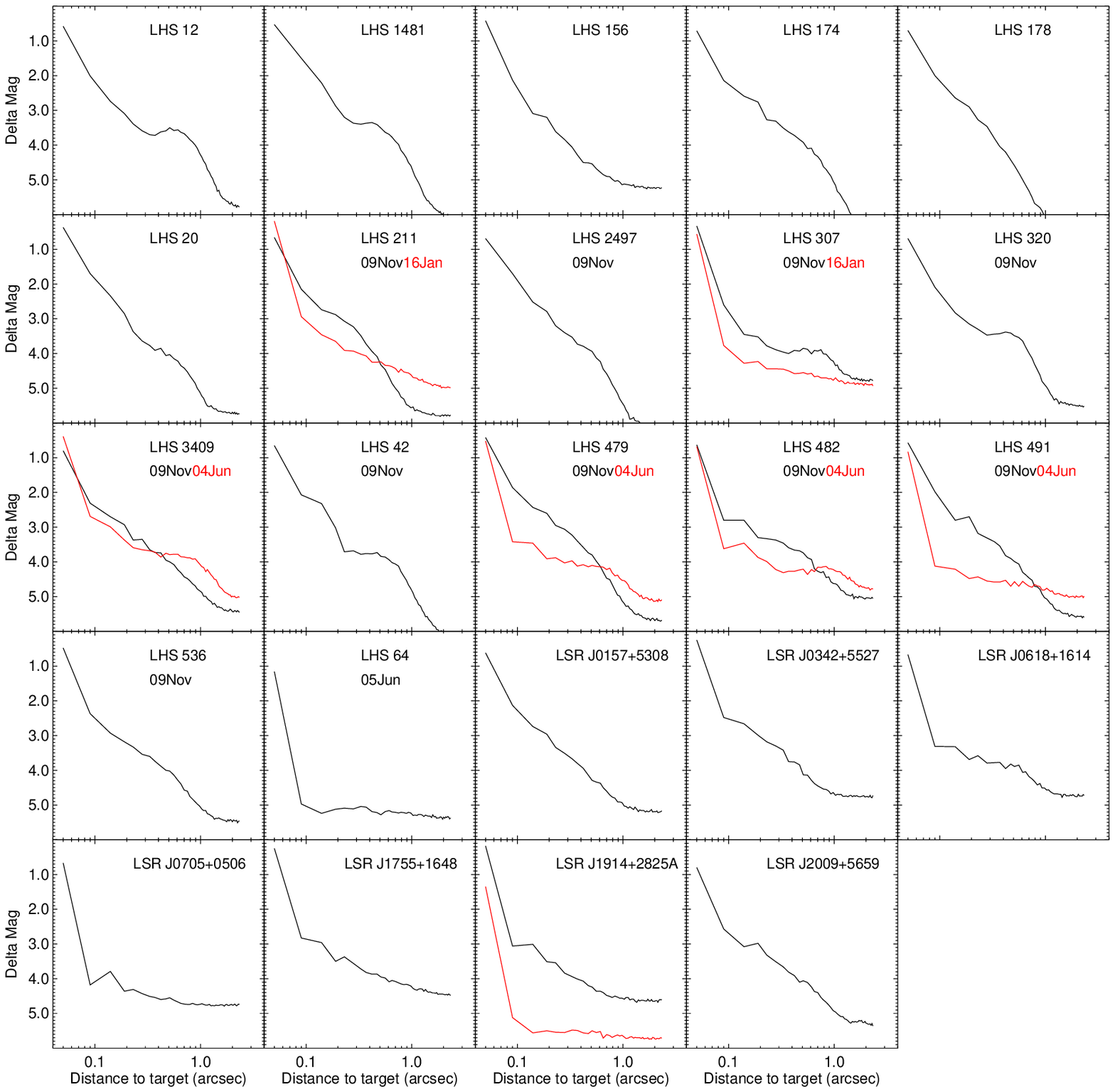}
   \caption{Difference in magnitude (Delta Mag) between the subdwarf
targets and the sky as a function of separation from the primary
for the best 10\% of all AstraLux images.
}
   \label{fig_subdw_CAHA:DeltaMag_sdM}
\end{figure*}

\begin{figure*}
   \centering
   \includegraphics[width=0.8\linewidth]{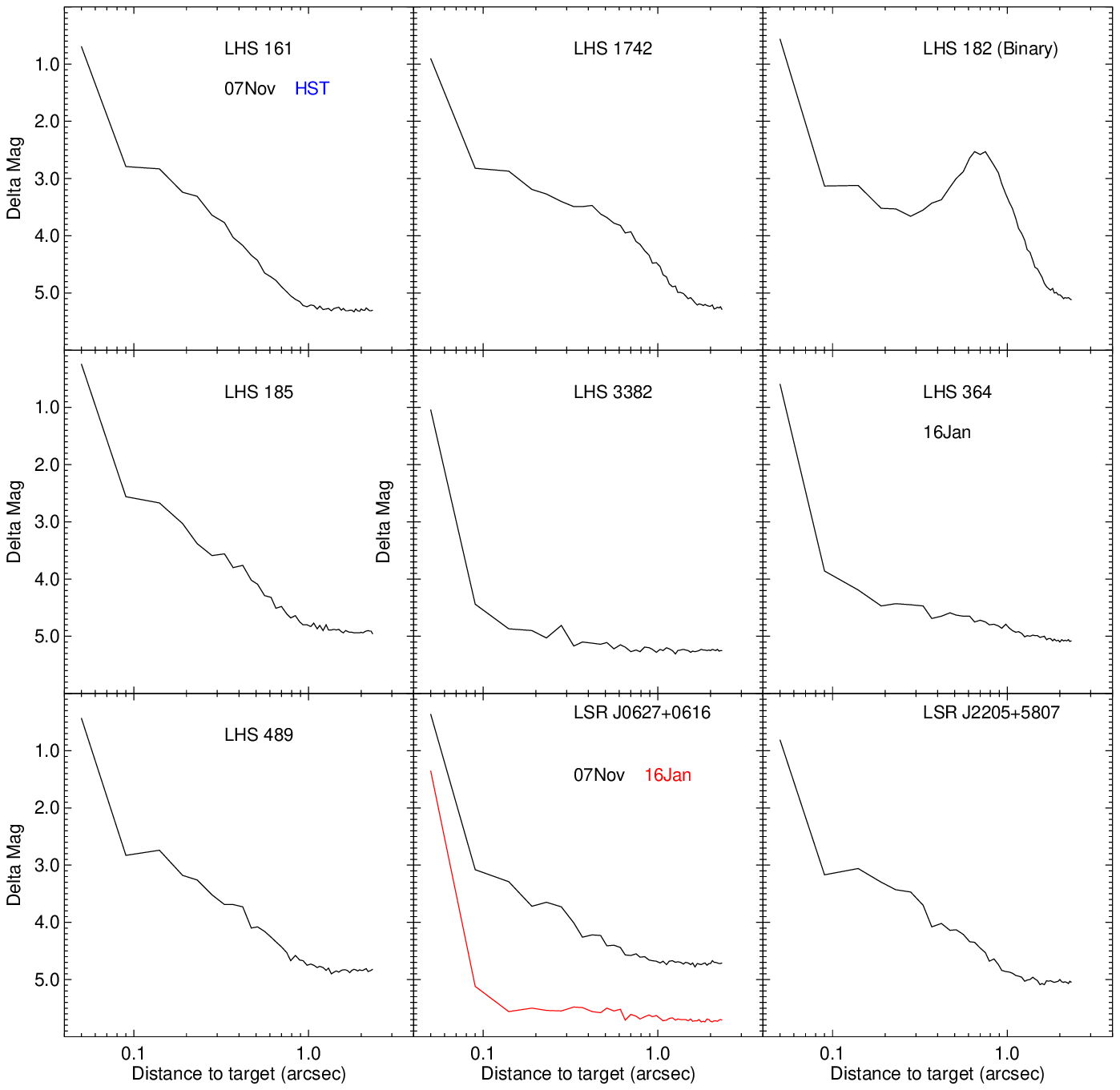}
   \caption{Difference in magnitude (Delta Mag) between the extreme
subdwarf targets and the sky as a function of separation from the primary
for the best 10\% of all AstraLux images.
}
   \label{fig_subdw_CAHA:DeltaMag_esdM}
\end{figure*}

The selection of the best 10\% of the images provides diffraction-limited 
images for 2-m class telescopes i.e., full-width-half-maximum (FWHM) of 
the order of 0.1 arcsec at red (0.6--0.9 microns) optical wavelengths.
We have measured the number of counts from the
sky at different radii from the object to estimate the depth achieved
on each image as a function of the distance from the target ($\Delta$Mag).
We incremented the radii by 2 pixels and assumed a pixel scale of
0.0233 pixel resulting from the Drizzled algorithm applied to the
AstraLux images. We typically achieved a $\Delta$Mag of $\sim$2 magnitudes 
at $\sim$0.1 arcsec from the target and 4--5 magnitudes at $\sim$1.0 arcsec
(Table \ref{tab_subdw_CAHA:sample_observed}).
We are sensitive to separations as large as 300 AU at a distance of
50 pc, the size of the AstraLux detector (6 arcsec in each direction).
The results are shown in Figs.\ \ref{fig_subdw_CAHA:DeltaMag_sdM} and
Fig.\ \ref{fig_subdw_CAHA:DeltaMag_esdM} for the 24 sdM and 9 esdM,
respectively. Figure \ref{fig_subdw_CAHA:DeltaMag_esdM} includes
the new binary system, LHS\,182, resolved with AstraLux.
We give the $\Delta$Mag at 0.47 and 0.98 arcsec
(corresponding roughly to 20 and 40 pixels, respectively) from the target
in Table \ref{tab_subdw_CAHA:sample_observed}. Note that several objects
were observed on two different nights and we list the
best $\Delta$Mag for them in Table \ref{tab_subdw_CAHA:sample_observed}
but we show the depth achieved as a function of the separation to the
primary for both nights (Figs.\ \ref{fig_subdw_CAHA:DeltaMag_sdM} \&
\ref{fig_subdw_CAHA:DeltaMag_esdM}).

To assess the depth and resolution achieved with the AstraLux images,
we have compared our sensitivity curves with {\it{HST}} observations of
seven subdwarfs observed by \citet{riaz08a} and common to our sample.
The depth within 0.5 arcsec of the target is greater in {\it{HST}}
observations than in our ground-based images (typically 6 vs 2 mag).
The depth by the {\it{HST}} at larger distances is typically three
magnitudes better than for our AstraLux data. The range of resolution
and depth achieved by the Lucky Imaging technique lie within the
sensitivity figures of the {\it{HST}}, implying that we can add
the sample of subdwarfs targeted by \citet{riaz08a} to our sample
to discuss the binary properties of low-metallicity M dwarfs.

We have found objects close to our targets on the AstraLux images in three
cases but the companionship is unlikely in two cases, LHS\,489 and
LSR J2009$+$5659, due to the large proper motion of the objects
(Fig.\ \ref{fig_subdw_CAHA:images_motion}). In the former case,
LHS\,489 is moving southwards
($\mu_{\alpha}cos\delta$=$-$0.053, $\mu_{\delta}$=$-$1.219 arcsec/yr) towards
two stars oriented north-south and separated by about four arcsec.
The expected position of LHS\,489 at the epoch of the AstraLux
observations based on its proper motion is consistent with a
separation of $\sim$2 arcsec with star \#1
(top panel in Fig.\ \ref{fig_subdw_CAHA:images_motion}).
We have repeated a similar analysis using the proper motion of
LSR J2009$+$5659 ($\mu_{\alpha}cos\delta$=$+$0.431; $\mu_{\delta}$=$+$0.700 arcsec/yr)
and determined that it should now be located north-east of
2MASS J200934.16$+$565930.2 (source \# 1 in bottom panels of
Fig.\ \ref{fig_subdw_CAHA:images_motion}) during our November observing
run, consistent with the position on the AstraLux images.

Figure \ref{fig_subdw_CAHA:images_motion} displays the image showing the 
motions of LHS\,489 (top) and LSR J2009$+$5659 over $\sim$55 years on 
POSS I (year 1953; left), POSS II (year 1990; middle), and AstraLux 
images (SDSSz; November 2008; right). The size of the POSS images is 
two arcmin across whereas the AstraLux images are 12 arcsec wide. The 
objects marked with numbers and detected on the AstraLux images are 
unlikely to be physically associated with the targets (circled) because 
they were already present on the photographic plates and appear closer 
to the target due to the large motions of LHS\,489 (1.22 arcsec/yr) and 
LSR J2009$+$5659 (0.822 arcsec/yr).

\begin{figure}
    \centering
    \includegraphics[width=0.49\linewidth]{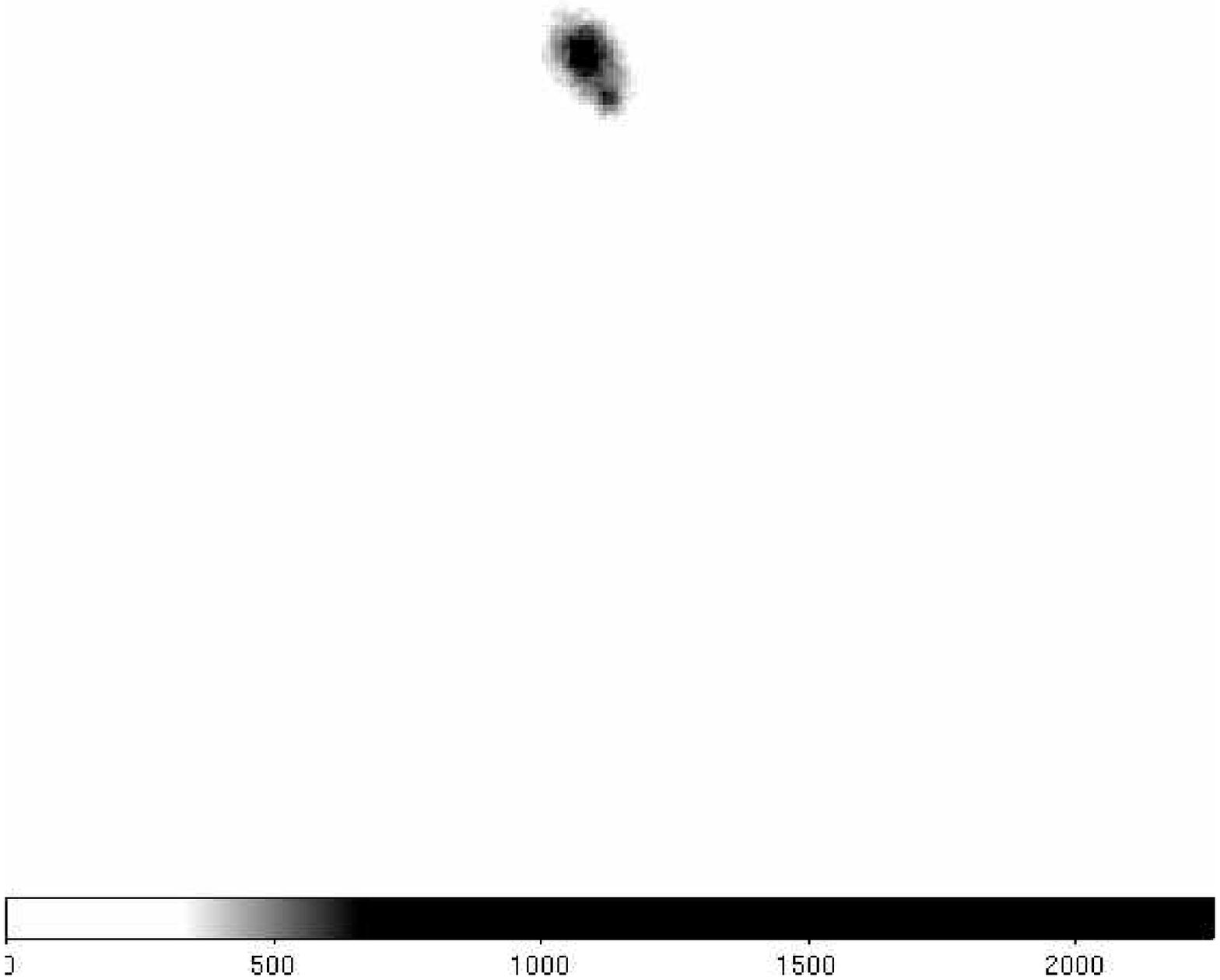}
    \includegraphics[width=0.49\linewidth]{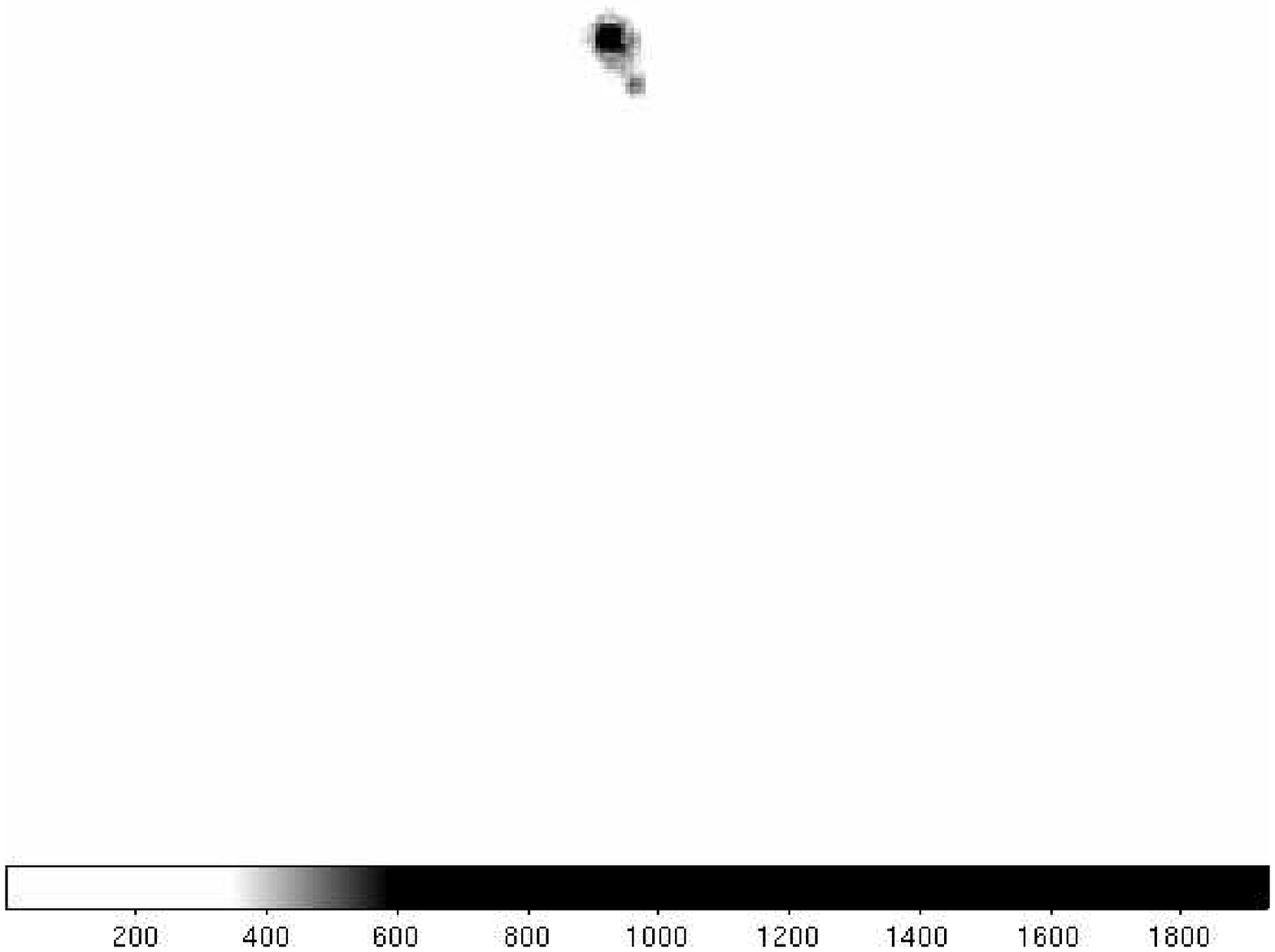}
    \caption{AstraLux images of the LHS\,1589AB system
    in the SDSS $i$ (left) and $z$ (right) filters.
 }
    \label{fig_subdw_CAHA:image_LHS1589}
 \end{figure}

We verify the astrometric companionship of
LHS\,1589A\&B discovered with the Lick adaptive optics laser guide star
system by \citet{lepine07b}. The epoch difference between the Lick
(16/17 September 2006) and the latest AstraLux (November 2008) observations
is greater than two years. We have measured a separation of 11 pixels
or 0.25$\pm$0.02 arcsec (Fig.\ \ref{fig_subdw_CAHA:image_LHS1589}).
Figure \ref{fig_subdw_CAHA:image_LHS1589} displays the AstraLux images
of the LHS\,1589AB system \citep{lepine07b} in the SDSS $i$ (left) and 
$z$ (right) filters. The pixel scale is 23.3 mas and the field-of-view 
is about 3.5 arcsec aside with North up and East left. The separation of 
0.224$\pm$0.004 arcsec measured by \citet{lepine07b} is in agreement with
the separation on the AstraLux images (0.25$\pm$0.02), confirming the 
common proper motion of the system.

During this period, LHS\,1589A has moved by more than 1.6 arcsec, implying
that the separation should have increased significantly if the companion
was not associated with LHS\,1589A\footnote{We also observed LHS\,1589AB
in November 2007 and measured a separation of $\sim$0.2 arcsec} .
As a consequence, we confirm the LHS\,1589AB pair as a true metal-poor
([m/H]$\sim$--1.0) and low-mass (0.3 M$_{\odot}$) binary system
\citep{lepine07b}.

%
%
%

%
%
%

%
%
\begin{table*}
 \centering
  \caption{Subdwarfs and extreme subdwarfs observed at high resolution in 
  the $z$-band with AstraLux.
}
 \label{tab_subdw_CAHA:sample_observed}
 \begin{tabular}{@{\hspace{0mm}}l c c c c c c c c c c c c c c@{\hspace{0mm}}}
 \hline
Name           &  R.A. (J2000) & dec (J2000)  &   $I$ & $J$   &  PM   & SpT      & d     &  Date    &  T  & \#Exp    & Strehl   & Res & 20p & 42p \cr
\ldots{}       &  h:m:s        & d:':"        &  mag  & mag   &  "/yr & \ldots{} & pc    & \ldots{} &  ms & \ldots{} & \ldots{} & pix & mag & mag \cr
 \hline
LSR J0157   & 01:57:40.63 &  +53:08:13.2 &  13.3 & 12.01 &  0.64 & sdM3.5  &  45.0 &  08Nov       &  50 & 6000  & 14.1 & 2.88 & 3.6 &  4.2  \\
LHS\,12     & 02:02:52.16 &  +05:42:21.0 &  11.2 &  9.47 &  2.44 & sdM0.0  &  27.9 &  08Nov       &  15 & 10000 &  9.0 & 3.20 & 3.0 &  4.1  \\
LHS\,156    & 02:34:12.46 &  +17:45:50.5 &  12.9 & 11.43 &  1.18 & sdM3.0  &  36.1 &  08Nov       &  30 & 10000 & 12.9 & 2.83 & 4.5 &  5.1  \\
LHS\,1481   & 02:58:10.23 &  -12:53:05.8 &  99.9 &  8.95 &  0.62 & sdM3.0  &  10.5 &  08Nov       &  15 & 10000 &  8.9 & 4.00 & 3.4 &  4.6  \\
LHS\,174    & 03:30:44.82 &  +34:01:07.2 &  11.1 &  9.84 &  1.57 & sdM0.5  &  49.0 &  07Nov       &  30 & 10000 & 13.1 & 4.32 & 3.8 &  5.4  \\
LHS\,20     & 03:38:15.58 &  -11:29:10.3 &  99.9 &  9.63 &  0.88 & sdM2.5  &  15.4 &  08Nov       &  50 & 6000  & 10.0 & 3.44 & 3.0 &  4.1  \\
LHS\,178    & 03:42:29.45 &  +12:31:33.8 &  99.9 &  9.11 &  0.53 & sdM1.5  &  22.2 &  07Nov       &  15 & 20000 & 14.8 & 4.46 & 4.4 &  6.0  \\
LSR J0342   & 03:42:53.73 &  +55:27:30.4 &  13.8 & 12.88 &  0.50 & sdM0.0  & 150.0 &  08Nov       &  50 & 6000  & 15.6 & 2.48 & 3.4 &  4.6  \\
LHS\,211    & 05:48:00.19 &  +08:22:14.2 &  12.3 & 11.19 &  1.27 & sdM0.0  &  53.2 &  16Jan,09Nov &  30 & 10000 & 16.2 & 3.27 & 4.1 &  5.6  \\
LSR J0618   & 06:18:52.54 &  +16:14:56.0 &  12.9 & 12.74 &  0.65 & sdM2.0  &  85.0 &  16Jan,07Nov &  50 & 6000  & 14.0 & 2.32 & 3.8 &  4.5  \\
LSR J0705   & 07:05:48.77 &  +05:06:17.3 &  13.4 & 13.69 &  0.51 & sdM3.5  &  90.0 &  07Nov       &  50 & 10000 & 18.1 & 2.01 & 4.6 &  4.7  \\
LHS\,307    & 11:32:45.28 &  +43:59:44.5 &  13.3 & 12.25 &  1.15 & sdM0.5  &  54.6 &  16Jan,09Nov &  50 & 6000  & 12.0 & 2.41 & 3.9 &  4.2  \\
LHS\,42     & 11:40:20.26 &  +67:15:35.0 &  11.4 & 09.41 &  3.17 & sdM0.0  &  30.6 &  09Nov       &  30 & 10000 & 10.7 & 3.46 & 3.7 &  4.8  \\
LHS\,2497   & 12:02:18.19 &  +28:35:14.3 &  10.2 & 09.13 &  0.79 & sdM3.5  &  20.3 &  09Nov       &  15 & 10000 & 11.2 & 4.43 & 3.9 &  5.4  \\
LHS\,320    & 12:02:33.65 &  +08:25:50.6 &  11.6 & 10.74 &  1.18 & sdM2.0  &  38.5 &  09Nov       &  30 & 10000 & 11.4 & 3.33 & 3.4 &  4.9  \\
LSR J1755   & 17:55:32.76 &  +16:48:59.0 &  12.5 & 11.35 &  0.99 & sdM3.5  &  28.0 &  04Jun,09Nov &  50 & 6000  &  8.8 & 2.28 & 3.9 &  4.2  \\
LHS\,3409   & 18:45:52.37 &  +52:27:40.0 &  12.9 & 10.97 &  0.85 & sdM4.5  &  20.0 &  04Jun,09Nov &  50 & 6000  & 10.7 & 3.23 & 3.9 &  4.8  \\
LSR J1914A  & 19:14:05.50 &  +28:25:52.3 &  13.6 & 13.53 &  0.53 & sdM0.0  & 200.0 &  07Nov       &  50 & 5000  & 16.9 & 2.39 & 4.2 &  4.6  \\
LHS\,479    & 19:46:48.60 &  +12:04:58.1 &  12.9 & 11.20 &  1.48 & sdM1.0  &  44.6 &  04Jun,09Nov &  50 & 6000  & 12.3 & 3.63 & 3.0 &  4.1  \\
LHS\,482    & 20:05:02.28 &  +54:26:03.8 &  12.3 & 08.83 &  1.47 & sdM1.5  &  53.2 &  04Jun,09Nov &  50 & 10000 & 14.2 & 2.93 & 3.7 &  5.1  \\
LSR J2009   & 20:09:33.82 &  +56:59:25.8 &  13.1 & 11.86 &  0.82 & sdM2.0  &  55.0 &  05Jun,09Nov &  50 & 6000  & 15.6 & 3.03 & 4.1 &  4.9  \\
LHS\,491    & 20:27:29.06 &  +35:59:24.5 &  12.4 & 11.60 &  1.31 & sdM1.5  &  47.4 &  04Jun,09Nov & 50 &  6000  & 14.5 & 3.27 & 3.9 &  5.0  \\
LHS\,64     & 21:07:55.43 &  +59:43:19.9 &  10.9 & 10.12 &  2.11 & sdM1.5  &  23.9 &  05Jun,09Nov &  15 & 10000 & 14.9 & 3.53 & 4.5 &  5.7  \\
LHS\,536    & 23:08:26.08 &  +31:40:24.0 &  12.1 & 11.71 &  1.52 & sdM0.5  &  44.0 &  09Nov       &  50 & 6000  & 13.6 & 3.14 & 4.0 &  5.0  \\
\hline
\hline
LHS\,161    & 02:52:45.51 &  +01:55:50.6 &  12.7 & 11.71 &  1.55 & esdM2.0 &  38.5 &  07Nov       &  30 & 10000 & 14.5 & 2.79 & 4.3 &  5.2  \\
LHS\,182    & 03:50:13.89 &  +43:25:40.5 &  12.6 & 11.10 &  1.44 & esdM0.0 &  42.7 &  07Nov       &  50 & 6000$^{a}$ &  9.7 & 2.48 & 3.4 &  3.6  \\
LHS\,185    & 04:01:36.60 &  +18:43:39.9 &  13.0 & 14.63 &  1.17 & esdM0.5 &  59.9 &  07Nov       &  50 & 6000  & 13.4 & 2.63 & 4.0 &  4.8  \\
LHS\,1742   & 05:10:31.41 &  +31:17:35.4 &  14.3 & 99.99 &  0.83 & esdM5.5 &  11.2 &  07Nov       &  50 & 6000  & 10.5 & 3.01 & 3.6 &  4.5  \\
LSR J0627   & 06:27:33.31 &  +06:16:58.8 &  99.9 & 13.29 &  1.02 & esdM1.5 &  80.0 &  16Jan,07Nov &  50 & 6000  & 16.9 & 2.15 & 4.2 &  4.7  \\
LHS\,364    & 14:06:55.54 &  +38:36:57.8 &  12.9 & 11.47 &  1.05 & esdM1.5 &  26.7 &  16Jan       &  50 & 6000  & 10.7 & 4.02 & 4.6 &  4.8  \\
LHS\,3382   & 18:21:52.95 &  +77:09:30.1 &  14.0 & 13.86 &  0.77 & esdM2.5 &  96.2 &  09Nov       &  50 & 6000  & 18.8 & 1.94 & 5.1 &  5.3  \\
LHS\,489    & 20:19:04.58 &  +12:35:04.1 &  13.9 & 12.53 &  1.24 & esdM0.0 &  52.9 &  09Nov       &  50 & 6000  & 14.6 & 2.58 & 4.1 &  4.8  \\
LSR J2205   & 22:05:32.78 &  +58:07:26.8 &  99.9 & 12.12 &  0.54 & esdM1.0 &  70.0 &  05Jun,09Nov &  50 & 6000$^{b}$ & 18.1 & 2.78 & 4.1 &  4.9  \\
\hline
 \end{tabular}
\end{table*}

%
%
\begin{figure}
    \centering
    \includegraphics[width=0.49\linewidth]{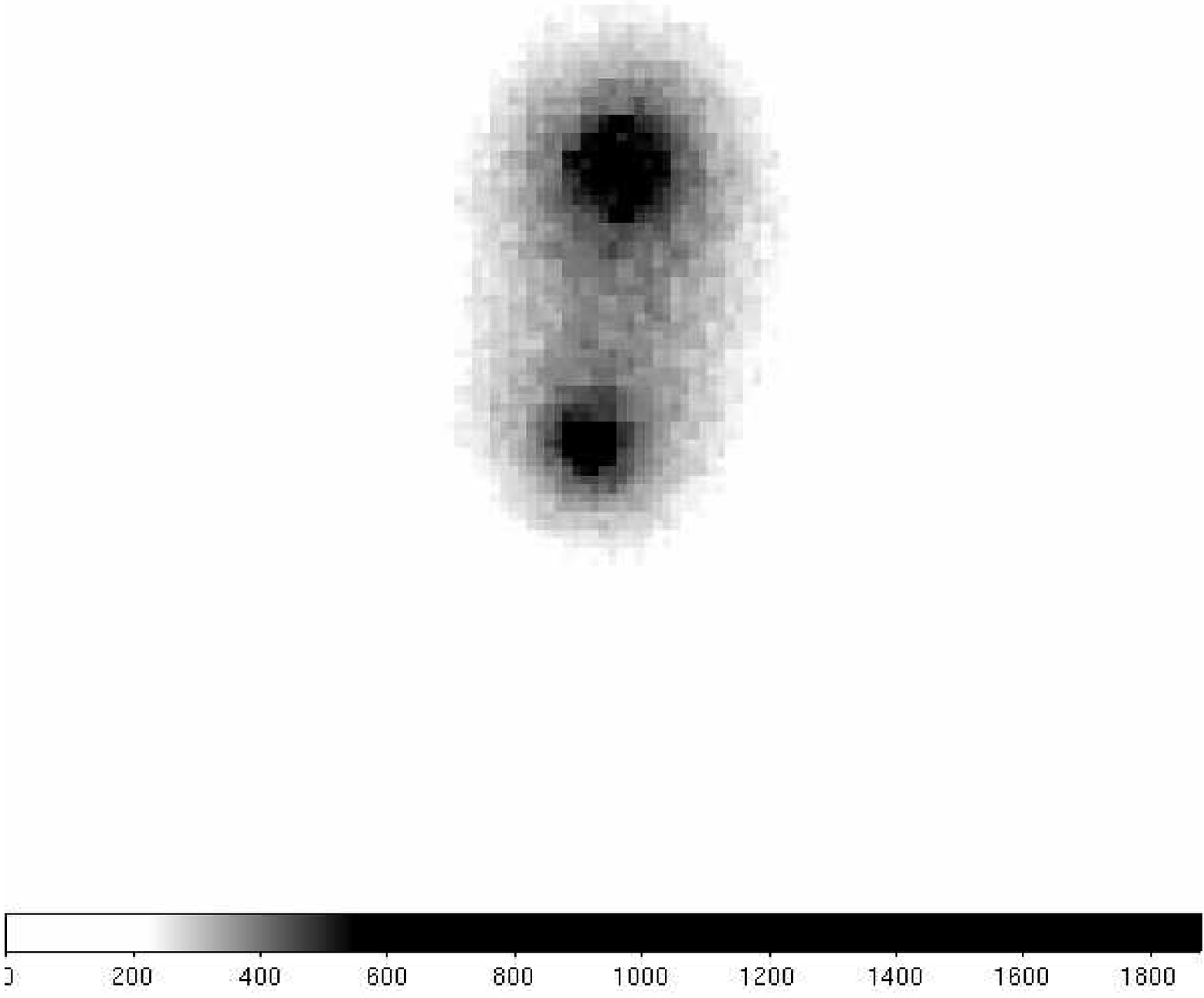}
    \includegraphics[width=0.49\linewidth]{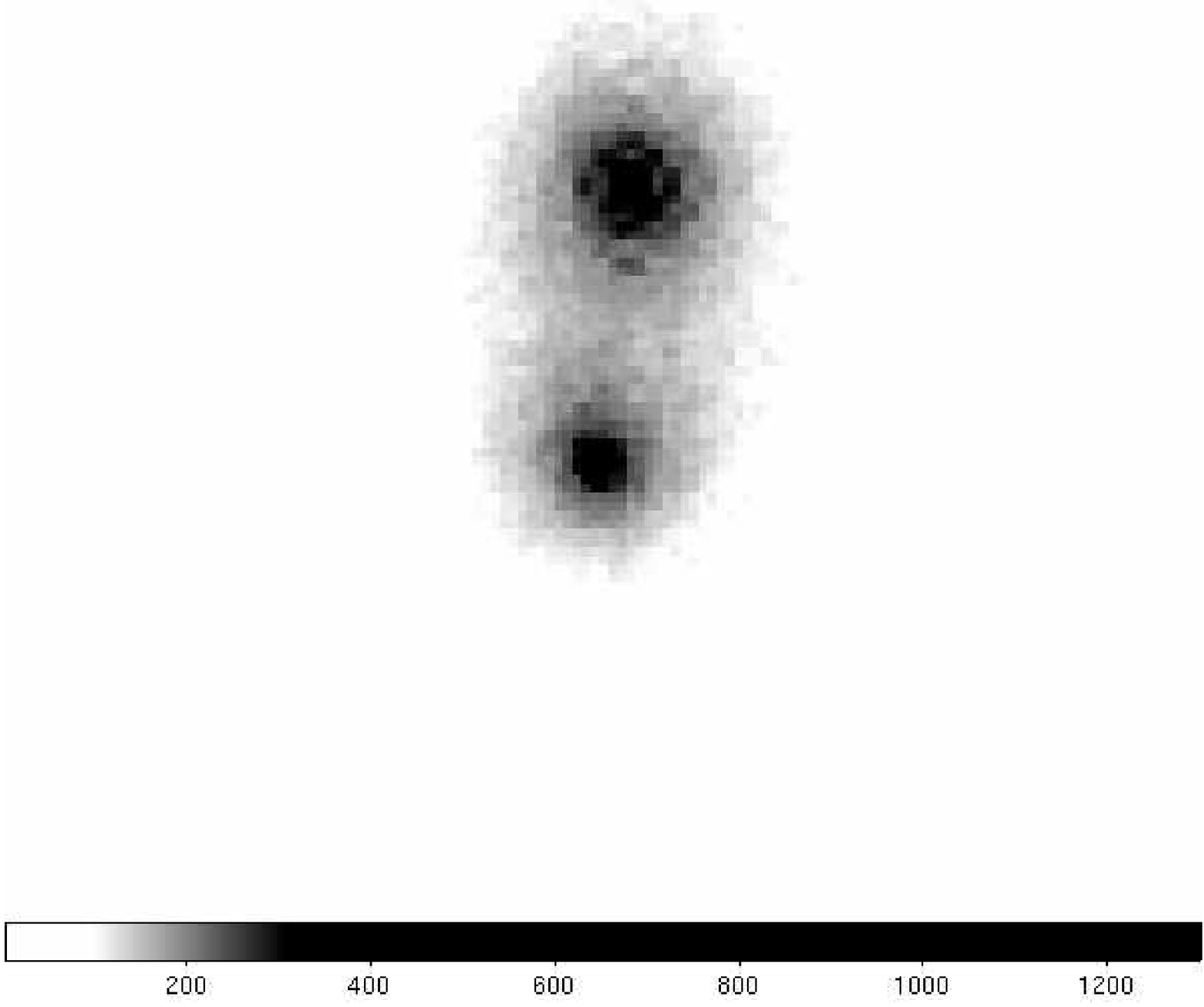}
    \caption{AstraLux Images of LHS\,182 in the SDSS $i$ (left) and $z$
    (right) filters.
 }
    \label{fig_subdw_CAHA:image_LHS182}
 \end{figure}
%

%
%
%

%
%
\subsection{LHS\,182: an extreme subdwarf binary}
\label{subdw_CAHA:case_LHS182}

Observations in November 2008 of LHS\,182 
\citep[esdM0; d = 42.7 pc;][]{harrigton80} revealed a close companion at 
$\sim$0.7 arcsec (Fig.\ \ref{fig_subdw_CAHA:image_LHS182}), an object detected 
neither on photographic plates nor in 2MASS \citep{cutri03} after taking into
account the proper motion of LHS\,182
\citep[$\mu_{\alpha}$=$-$0.374, $\mu_{\delta}$=$-$1.391 arcsec/yr;][]{lepine05d} as we did for e.g.\ LHS\,489\@.
Figure \ref{fig_subdw_CAHA:image_LHS182} shows the SDSS $i$ (left)
and $z$ (right) filters mounted on AstraLux on the Calar Alto 2.2-m 
telescope. The pixel scale is 23.3 mas and field-of-view is about 
three arcsec aside with North up and East left.

The closest objects detected in previous surveys are located at a few
arcsec from the position of LHS\,182 in November 2008\@. Therefore,
we consider this system as a true physical binary. After re-processing
the raw images to remove the effect of the ``triple'' system (see details
in Sect.\ \ref{subdw_CAHA:Obs_DR}), we have measured a brightness ratio
of 0.1--0.2 mag in the $z$-band and a separation of 0.70$\pm$0.05 arcsec,
corresponding to a projected physical separation of about 30 AU at the
distance of LHS\,182\@. An image taken three months later (on 13 February
2009) with AstraLux confirms the common proper motion of the pair and
the separation of $\sim$0.7 arcsec. We also note that this pair was
reported by \citet{jao09a} and a similar separation was reported
(0.62 arcsec) from observations made in November 2005\@.

%
%
\section{Binary fraction of early-M subdwarfs}
\label{subdw_CAHA:BF}

In this section we place our results in a wider context and discuss the
possible role played by the metallicity in the binary properties of
low-mass stars.

We have detected one close companion to LHS\,182, an esdM0 at $\sim$43 pc,
from a sample of 33 metal-poor early-M dwarfs
(Table \ref{tab_subdw_CAHA:sample_observed}).
We have shown in Sect.\ \ref{subdw_CAHA:analysis} that we are able 
to resolve LHS\,1589AB into a binary system with a separation of 
$\sim$0.2 arcsec and a difference in $K$ of 0.52 mag \citep{lepine07b}, 
suggesting that the resolution obtained with AstraLux is comparable to 
high-resolution imaging surveys with adaptive optics and from space. 
From our AstraLux sample alone, we infer a binary fraction of
1/33 = 3$\pm$3\% (Poissonian errors) at separations wider than 5 AU 
around metal-poor early-M dwarfs (assuming a mean distance of 50 pc). 
This binary frequency is comparable to the fraction derived by 
\citet{riaz08a} from a sample of 28 objects (3.6$\pm$3.6\%; we have 
shown in Sect.\ \ref{subdw_CAHA:lhs1074} that the companion found around
LHS\,1074 does not share the same proper motion) and by \citet{lepine07b}
from 18 subdwarfs (5.6$\pm$5.6\%). We note that the only binary found 
by \citet{riaz08a} is not a subdwarf binary but a wide common proper 
motion pair composed of a subdwarf and a white dwarf 
\citep[LHS\,2139/2140;][]{luyten79}. Adding the 28 subdwarfs from 
\citet{riaz08a} to our sample (and taking into account the seven objects 
in common), the fraction of 0.1--0.4 M$_{\odot}$ subdwarf binaries is 
2/(33$+$28$-$7) = 3.7$\pm$2.6\% (1$\sigma$ confidence level) for the 
aforementioned separation range. The binary fraction of solar-metallicity 
M dwarfs with masses in the 0.13--0.6 M$_{\odot}$ range is 20--25\% 
\citep{fischer92,reid97a} for separations in the 6--300 AU range. 
Therefore, there is a significant difference (a factor of two with
a 3$\sigma$ confidence level)
between the binary frequency of metal-poor early-M dwarfs and
their solar counterparts in that separation range. Below we discuss 
the possible reasons for this discrepancy and the role that metallicity
seems to play at lower masses.
There are several alternatives to account for the deficit of binary
systems at low metallicity.

One possible explanation could reside in the difference in the composition
of the metal-poor and solar-metallicity parent clouds, leading to distinct
binary properties. However, metal-poor and solar-metallicity stars seem
to share the binary characteristics (frequency and orbital parameters)
over the entire separation range for masses above $\sim$1 M$_{\odot}$
\citep{partridge67,stryker85,allen00,latham02,zapatero04a}.
Surveys involving thousands of radial velocity measurements of 
hundredth of stars \citep{carney94} taken over several years 
\citep{stryker85,latham02} concluded that the frequency of 
spectroscopic binaries for both metal-poor and solar-metallicity 
G stars is around 15\% \citep{duquennoy91}, in agreement with the 
early study of \citet{partridge67}.
At very wide projected physical separation (a $>$ 25 AU), the picture
seems comparable with 15\% of subdwarfs exhibiting companions
\citep{allen00,zapatero04a}. At intermediate separations, the
frequency of metal-poor stars seems again in agreement with the
fraction of solar-metallicity G stars \citep{zinnecker04}.
As pointed out by \citet{latham02}, if the composition of the initial 
molecular cloud plays a role in setting the binary properties, it is
unlikely to be the main reason for the deficit of metal-poor
low-mass stars.

Nonetheless, metallicity may play a role early on in the formation of 
such multiple systems as appears to be the case for the formation 
of massive planets around solar-mass stars. This fact is supported by 
the higher number of planets orbiting metal-rich stars than 
solar-metallicity stars \citep{santos05a,bond06} coupled with the
non-detection of planets in 47 Tuc \citep[Fe/H = $-$0.7;][]{gilliland00}. 
The amount of metals in the molecular clouds would favour the presence 
of solid planetesimals on which dust accumulates to grow and ultimately 
form giant planets \citep{santos01a}. However, this process is still not 
very well understood.

Metal-poor stars tend to be older than solar-abundance stars and
have thus suffered more encounters with other stars. The large majority
may stem from globular clusters whose high 
density favour the disruption of binaries with separations larger 
than a few AU\@. Therefore, we are measuring the results of dynamical
interactions between stars in dense environments.
However, the tightest systems with the highest binding energy should be 
less affected by the process of ejection \citep{sterzik98}, pointing in
the direction of a higher binary frequency for solar-mass stars than
for M dwarfs. If this assumption is true, radial velocity surveys 
should find a higher frequency for spectroscopic binaries than our
results although it may remain lower than the fraction inferred 
for solar-metallicity M dwarfs. This effect can be compared with the
difference observed between the binary frequency of low-mass stars in 
the dense Orion cluster and the low-density Taurus star--forming region
\citep{koehler06}: they observed a factor of 3--5 difference between
both clusters. However, they found that the binary fraction is
roughly constant when going radially out from the centre, suggesting
that disruption is not the main effect. The most likely explanation
is the extremely dense environment dominating during the formation
process of massive open clusters like the Trapezium Cluster
\citep{kroupa01a}. This effect would be magnified for globular clusters
and could lead to the difference observed between metal-poor and 
solar-metallicity M dwarfs \citep[e.g., the lack of planets
in 47 Tuc;][]{gilliland00}.


Moreover, low-metallicity stars could populate the Milky Way as
a result of mergers with satellite galaxies.
For example, \citet{meza05} and \citet{abadi06} presented simulations
predicting that a large number of nearby metal-poor stars would
originate from a merger with a satellite galaxy like $\omega$ Cen.
They would then remain in the Galaxy after suffering significant
perturbations that would likely modify their primordial binary
properties, which are unknown. Besides, the environment in those
galaxies could be very different to what we know in our Galaxy.

Lastly, another potential explanation could be that the mass-luminosity
relation for low-mass stars and brown dwarfs cools down faster than
predicted by theoretical models and/or is steeper for a given age. This
interpretation, although less probable than the aforementioned proposals, 
cannot be discarded with current data.
 
%
%
\section{Conclusions}
\label{subdw_CAHA:conclusions}

We have presented high-resolution imaging for a sample of 33 M subdwarfs
located within 100 pc of the Sun. We have uncovered one close (projected
physical separation of $\sim$ 30 AU) companion to the M0 extreme subdwarf
LHS\,182\@. No companion was resolved around the other targets
at separations larger than 10 and 50 AU (at a distance of 50 pc) down to 
two and five magnitudes fainter than the primary, respectively.
Moreover, we do not confirm the common proper motion candidate of
LHS\,1074 discovered by \citet{riaz08a} on astrometric and photometric
grounds. Our results support previous surveys of M subdwarfs and suggests
that the binary frequency of M dwarfs is metallicity-dependent. 

To test whether or not subdwarfs are born in globular clusters and survive
the high densities present at the early formation stage, radial velocity 
surveys of a large sample of M subdwarfs and extreme subdwarfs should be 
conducted. The main drawback is the large amount of time required on 
8 to 10-m class telescopes to achieve good signal-to-noise ratios at those 
faint magnitudes. However, such a survey would shed light on the origin of
halo interlopers. Finally, many ultracool subdwarfs have been discovered 
recently \citep[e.g.,][]{lepine08b} and this rate will likely increase 
with current (e.g., UKIDSS) and upcoming large-scale surveys (Visible and 
Infrared Survey Telescope for Astronomy\footnote{http://www.vista.ac.uk/}; 
PanSTARS\footnote{http://pan-starrs.ifa.hawaii.edu/public/}; 
Large Synoptic Survey Telescope\footnote{http://www.lsst.org/}). Finding
metal-poor brown dwarfs and investigating their binary properties will
require large telescopes combined with Laser Guide Star facilities.

%
%
\begin{acknowledgements}
Support for this project has been provided by the Spanish
Ministry of Science via project AYA2007-67458\@.
Thanks to the Calar Alto staff for their support with the AstraLux
observations and Felix Hormuth for valuable information.
NL thanks Luisa Valdivielso and Florian Rodler for their assistance
with the Lucky Imaging observations.

This research has made use of the Simbad database, operated at
the Centre de Donn\'ees Astronomiques de Strasbourg (CDS), and
of NASA's Astrophysics Data System Bibliographic Services (ADS).

This publication makes use of data products from the Two Micron
All Sky Survey (2MASS), which is a joint project of the University
of Massachusetts and the Infrared Processing and Analysis
Center/California Institute of Technology, funded by the National
Aeronautics and Space Administration and the National Science Foundation.
\end{acknowledgements}

%
%
  \bibliographystyle{aa}
  \bibliography{../mnemonic,../biblio_old}

\end{document}